\documentclass{article}
\usepackage{
spconf,
amsmath,
graphicx,
amssymb,
algorithm,
algorithmic,
multirow,
booktabs,
tabularx,
footmisc,
xfrac,
enumitem,
float,
xurl,
}

\usepackage{hyperref}
\hypersetup{colorlinks,allcolors=black}

\title{Melody Harmonization Using Orderless NADE, Chord Balancing, and Blocked Gibbs Sampling}

\name{Chung-En Sun, Yi-Wei Chen, Hung-Shin Lee, Yen-Hsing Chen, Hsin-Min Wang}
\address{Institute of Information Science, Academia Sinica, Taiwan}
\begin{document}
\maketitle
\begin{abstract}
Coherence and interestingness are two criteria for evaluating the performance of melody harmonization, which aims to generate a chord progression from a symbolic melody. In this study, we apply the concept of orderless NADE, which takes the melody and its partially masked chord sequence as the input of the BiLSTM-based networks to learn the masked ground truth, to the training process. In addition, the class weights are used to compensate for some reasonable chord labels that are rarely seen in the training set. Consistent with the stochasticity in training, blocked Gibbs sampling with proper numbers of masking/generating loops is used in the inference phase to progressively trade the coherence of the generated chord sequence off against its interestingness. The experiments were conducted on a dataset of 18,005 melody/chord pairs. Our proposed model outperforms the state-of-the-art system MTHarmonizer in five of six different objective metrics based on chord/melody harmonicity and chord progression. The subjective test results with more than 100 participants also show the superiority of our model.
\end{abstract}


\section{Introduction}
\label{sec:introduction}

Automatic melody harmonization aims to build up a learning machine that can generate chord sequences to accompany a given melody \cite{Chuan2007,Simon2008}. In music, a chord is any harmonic set composed of three or more notes that are heard as if sounding simultaneously \cite{Makris2016}. However, melody harmonization is a highly subjective task that is difficult for machines to learn to imitate humans or fit human preferences. While some harmonization may be considered appropriate in some cases, it can be changed according to different contexts. 

Many approaches have been proposed for automatic melody harmonization \cite{Makris2016}, such as hidden Markov models (HMMs) \cite{Paiement2006,Tsushima2017,Tsushima2018} and genetic algorithm (GA)-based methods \cite{Kitahara2018}. Recently, with the prevalence of deep learning models, some deep learning methods have emerged to deal with the melody harmonization problem \cite{Briot2017}. Lim \textit{et al.} \cite{Lim2017} first proposed a model composed of bidirectional long short-term memory (BiLSTM) layers \cite{Hochreiter1997}. The model predicts a chord for each bar from 24 major and minor chords. Based on the same neural structure, Yeh \textit{et al.} extended the number of chords to 48 by considering major, minor, augment and diminish chords. In addition, they integrated information of chord functions \cite{Chen2018} into the loss function to help chord label prediction, and predicted a chord every half bar \cite{Yeh2020}. Experiments showed that their model, called MTHarmonizer, could generate more creative and elaborated chord progressions than the previous model. Not only the expanded diversity of chord labels, the reason also lies in that MTHarmonizer can handle the transition between tonic (T), subdominant (S), and dominant (D) chords due to the use of chord functions.

However, there are several drawbacks in the above models. For example, Lim's model overuses common chords and has an incorrect phrasing problem \cite{Yeh2020}. Yeh's model tried to deal with these problems, but still could not appropriately assign altered chords through chord function prediction, because the chord function (TSD) easily loses its function when a secondary or borrowed chord event occurs.

In this study, we propose a BiLSTM-based model stemming from the structure of orderless NADE \cite{Uria2014,Uria2016}, which takes the melody and its partially masked chord sequence as the input to learn the masked ground truth. Instead of chord functions, class weights are used to compensate for some reasonable chord labels that are rarely seen in the training set. In addition, we perform blocked Gibbs sampling during inference \cite{Yao2014}, similar to CoCoNet used in Google's Bach Doodle \cite{Huang2017,Huang2019}. As a result, our model can predict 96 chords, including major, minor, augment, diminish, suspend, major7, minor7, and dominant7. Through the chord balancing process in training, we finally achieve a generative model that can produce rich but still reasonable chord progressions. Such interesting progressions with sufficient tension have a close quality to that of human-composed progressions.

\section{Proposed model}
\label{sec:proposed model}

\subsection{Neural Structure}
\begin{figure}[!t]
\centering
\includegraphics[width=0.48\textwidth]{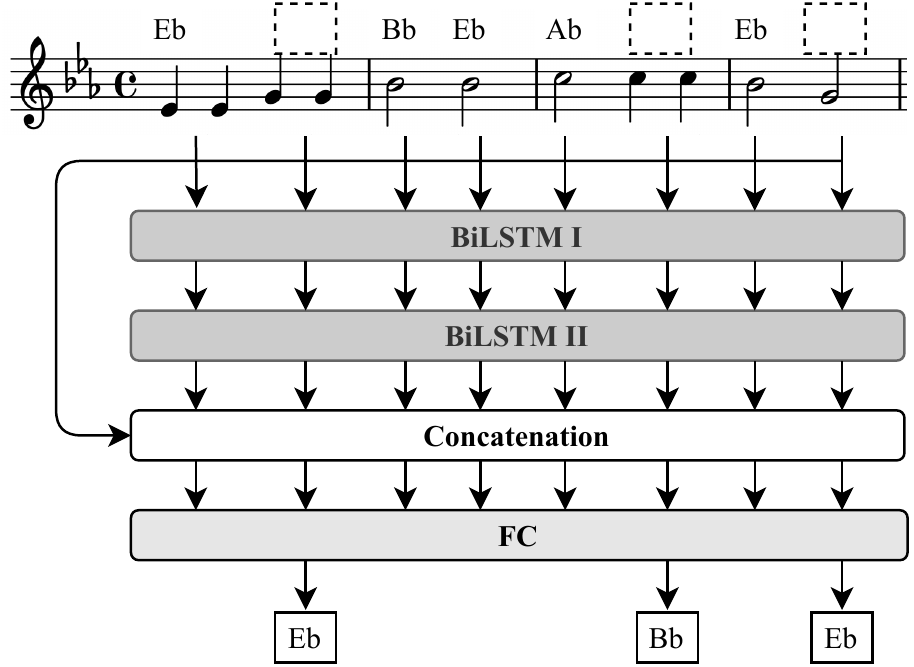}
\vspace{-20pt}
\caption{Proposed neural structure for melody harmonization. The input of each hidden state of BiLSTM I is a melody sequence within half a bar, concatenated with a chord that may be masked (dashed boxes) and the mask itself. The output of BiLSTM II is concatenated with the input to form a residual-like network \cite{He2016}, and then fed into a fully-connected (FC) layer for final prediction.}
\label{fig:model}
\vspace{-10pt}  
\end{figure}

Similar to Lim's \cite{Lim2017} and Yeh's \cite{Yeh2020} models, the neural structure of our model is mainly composed of two BiLSTM layers, followed by a fully connected layer, as shown in Fig. \ref{fig:model}. In the training phase, dropout with a probability of 0.2 is applied to each hidden layer \cite{Srivastava2014}, and mini-batch gradient descent is used to minimize the categorical cross entropy \cite{Ruder2016}. We use Adam as the optimizer with a learning rate of 0.005 and early stopping at the 10-th epoch \cite{Kingma2015}.

However, we employ the technique of orderless NADE to train our model in order to approximate the complex joint probability between melody and chords as well as chords and chords \cite{Uria2014,Uria2016}. Given any arbitrary context $C$, an orderless NADE model offers direct access to all distributions of the form $p_\theta({x_i}|{x_C}),i\notin C$, where $\theta$ denotes the model parameters and $x_i$ denotes the $i$-th variable that is not in the set of contextual variables $x_C$ that are already known. We can use it to complete any arbitrarily partial chord progression.

The training procedure is as follows. First, there are three tensors at the input, including a one-hot encoding tensor $x_\textnormal{melody} \in \{0,1\}^{T\times |P|}$, a one-hot encoding tensor $x_\textnormal{chord} \in \{0,1\}^{T\times |S|}$, and a binary mask $x_\textnormal{mask} \in \{0,1\}^{T\times 1}$. $T$ denotes the number of time steps. The number of pitches $|P|$, i.e., the number of notes in the chromatic scale, is set to 12. The number of chords $|S|$ is set to 96. $x_\textnormal{mask}$ is used to mask some parts of the $x_\textnormal{chord}$ and provides the model with information about which parts are masked:
\vspace{-5pt}
\begin{equation}
x_{\textnormal{chord}_C} = x_\textnormal{chord} \circ x_\textnormal{mask},
\vspace{-5pt}
\end{equation}
where $x_{\textnormal{chord}_C}= \{x_{\textnormal{chord}_t}|t \in C\}$ is a strict subset of $x_\textnormal{chord}$ based on context $C$, and $\circ$ denotes the Hadamard product.

Therefore, the model is only provided some part of $x_\textnormal{chord}$. The positions of $x_\textnormal{chord}$ to be masked are randomly selected, and the masking ratio follows a uniform distribution between 0 and 1. The three tensors are further concatenated to form the model input $x_\textnormal{input}$ as follows:
\vspace{-5pt}
\begin{equation}
x_\textnormal{input} = \mathrm{concat}(x_\textnormal{melody}, x_{\textnormal{chord}_C}, x_\textnormal{mask})
\vspace{-5pt}
\end{equation}

Finally, the task of our model is to reconstruct $x_{\textnormal{chord}_{\neg C}}$, the complement of $x_{\textnormal{chord}_C}$, from $x_\textnormal{input}$.

\subsection{Loss Function}
\label{sec:Loss Function}

The loss function with negative log-likelihood is expressed by
\vspace{-5pt}
\begin{equation}
\begin{split}
&\mathcal{L}(x_\textnormal{melody},x_\textnormal{chord};C,\theta ) =\\
&-\sum_{t\notin C} \sum_{s}w_{s}x_{\textnormal{chord}_{t,s}}\log p_{\theta}(x_{\textnormal{chord}_{t,s}}|x_\textnormal{melody},x_{\textnormal{chord}_C},C)
\end{split}
\vspace{-5pt}
\end{equation}
where $w_s$ is the weight of chord $s$ \cite{Cui2019}, $x_{\textnormal{chord}_{t,s}}$ is the $s$-th element of the one-hot vector $x_{\textnormal{chord}_t}$, $C$ is the context given, and $\theta$ denotes the model parameters. Chord weighting can greatly solve the problem of excessive use of common chords in Lim's model \cite{Lim2017} like Yeh's chord function \cite{Yeh2020}. However, our model can further manage to assign appropriate altered chords when the melody requires a secondary or borrowed chord to elaborate the whole piece, but Yeh's model \cite{Yeh2020} cannot. The weight matrix is derived by
\vspace{-5pt}
\begin{equation}
w_s = |S| \times {\dfrac{\sfrac{1}{(1000+n_s)}}{\sum\limits_{s'}{\sfrac{1}{(1000+n_{s'})}}}}
\vspace{-5pt}
\end{equation}
where $n_s$ is the number of times chord $s$ appears in the training data, and $n_s$ is added by 1000 for count smoothing. This process is also called inverse class frequency weighting.

The model is trained by minimizing
\vspace{-5pt}
\begin{equation}
\mathbb{E}_{x_\textnormal{chord}\sim p(x_\textnormal{chord})}\mathbb{E}_{C\sim p(C)}\dfrac{1}{|\neg C|}\mathcal{L}(x_\textnormal{melody},x_\textnormal{chord};C,\theta).
\vspace{-5pt}
\end{equation}
The expectations are estimated by sampling $x_\textnormal{chord}$ from the training set and selecting a single context $C$ per sample.

\subsection{Inference}
\label{sec:Inference}
The original orderless NADE model uses ancestral sampling in the inference stage \cite{Uria2014,Uria2016}, which samples one variable at a time in any order. Therefore, we can randomly choose an order, and then sample all the variables according to that order. However, this process may cause errors to propagate through that specific order. Furthermore, it is shown in \cite{Huang2017} that some $p_\theta(x_{\textnormal{chord}_t}|x_C,C)$, $t\notin C$, may be poorly modeled, especially when $C$ is small.

Therefore, we use blocked Gibbs sampling \cite{Yao2014}, a similar sampling procedure in \cite{Huang2017,Huang2019}. The blocked Gibbs sampling is an orderless sampling process that perfectly matches the training process of our model. There is no need to select an order first; instead, we randomly choose the blocks to be masked and ask the model to reconstruct them in each iteration. There is an annealing probability to determine the proportion of the variables that remain unchanged. At first, all the variables will be erased and sampled again. As the iteration goes on, more and more variables will be kept unchanged. Eventually, the output will converge and the joint probability can be estimated elaborately. Algorithm 1 elaborates our sampling process in the inference stage. The number of iterations $n$ is set to the average length of chord sequences, $P_\mathrm{max}$ is set to 1, and $P_\mathrm{min}$ is set to 0.05.

\renewcommand{\algorithmicrequire}{\textbf{Input:}}
\renewcommand{\algorithmicensure}{\textbf{Output:}}
\algsetup{linenosize=\normalsize,linenodelimiter=.}

\begin{algorithm}[t]
\caption{Blocked Gibbs Sampling}
\begin{algorithmic}[1]
\REQUIRE A melody $x_\textnormal{melody}$.
\ENSURE A chord sequence $y_\textnormal{chord}$.
\STATE $x_\textnormal{chord} \leftarrow \mathbf{0}$ and $x_\textnormal{mask} \leftarrow \mathbf{1}$
\STATE $x \leftarrow \mathrm{concat}(x_\textnormal{melody}, x_\textnormal{chord}, x_\textnormal{mask})$
\STATE $y_\textnormal{chord} \leftarrow \mathrm{model}(x)$
\STATE for {$i=0,1 \dots n$}, \\
    \hskip\algorithmicindent (a) $\alpha \leftarrow P_\mathrm{min}+\dfrac{(P_\mathrm{max}-P_\mathrm{min})\times i}{n}$ \\
    \hskip\algorithmicindent (b) $x_\textnormal{mask} \leftarrow \mathrm{create\_random\_mask}(\alpha)$ \\
    \hskip\algorithmicindent (c) $y_\textnormal{chord} \leftarrow y_\textnormal{chord}\circ x_\textnormal{mask}$ \\
    \hskip\algorithmicindent (d) $x \leftarrow \mathrm{concat}(x_\textnormal{melody}, y_\textnormal{chord}, x_\textnormal{mask})$ \\
    \hskip\algorithmicindent (e) $y_\textnormal{chord\_new} \leftarrow \mathrm{model}(x)$ \\
    \hskip\algorithmicindent (f) $y_\textnormal{chord} \leftarrow \mathrm{where}(x_\textnormal{mask}\textnormal{ equals 1}, y_\textnormal{chord\_new}, y_\textnormal{chord})$ \\
\end{algorithmic}
\end{algorithm}

\section{Experiments and Evaluations}
\label{sec:Experiments and Evaluations}

\subsection{Dataset}
\label{sec:Dataset}
The Hooktheory Lead Sheet Dataset (HLSD) \cite{Anderson} is used in this study. The dataset is collected through a community-based platform for users to upload lead sheets. It contains high-quality, human-arranged melodies with chord progressions and other information. After keeping song tracks containing melody/chord pairs, we constructed a dataset of 18,005 samples for this study. We divided the dataset into a training set containing 17,505 samples and a test set containing 500 samples. The dataset contains chords that are beyond our chord space, such as 9th, 11th, 13th, half-diminished, and slashed chords. We transferred these chords into one of the 96 chords according to their chord functions. The slashed chords were returned to their root positions. All the pieces were transferred into either C major or c minor. In addition, by sampling chords at the beginning and middle of a bar, we made the chords in the dataset last at least half a bar.

\subsection{Metrics}
\label{sec:metrics}

For objective evaluation, we use six different objective metrics introduced in \cite{Yeh2020}.
\vspace{-3pt}
\begin{itemize}[leftmargin=*]
\setlength\itemsep{-0.3em}
\item {\bf{Chord histogram entropy (CHE):}}
The entropy of the histogram of chord space.
\item {\bf{Chord coverage (CC):}}
The number of chords that are used in a chord sequence.
\item {\bf{Chord tonal distance (CTD):}}
The tonal distance proposed in \cite{Harte2006} for measuring the distance between two chords.
\item {\bf{Chord tone to non-chord tone ratio (CTnCTR):}}
The ratio of the number of chord tones to non-chord tones.  
\item {\bf{Pitch consonance score (PCS):}}
The consonance score that is computed through every note in a given chord.
\item {\bf{Melody-chord tonal distance (MCTD):}}
The tonal distance between a note and a chord in 6-D feature vectors.
\end{itemize}
\vspace{-3pt}
The first three metrics measure the quality of the chord progression, and the others measure the coherence between the melody and chords. For details about the metrics, see \cite{Yeh2020}.

For subjective evaluation, we randomly selected 50 melodies from our test set. Each human subject needs to evaluate the chord progressions of five melodies. For each melody, the subjects listen to the melody without chords first, and then three different harmonizations composed by humans, MTHarmonizer, and our model in random order.

We design four criteria to obtain feedback from subjects:
\vspace{-3pt}
\begin{itemize}[leftmargin=*]
\setlength\itemsep{-0.3em}
\item {\bf{Coherence:}}
how well the chord progression matches with the melody in terms of harmonicity and phrasing.
\item {\bf{Chord progression:}}
how reasonable or smooth the chord progression is on its own, independent from the melody.
\item {\bf{Interestingness:}}
how surprising or innovative the chord progression is on its own, independent from the melody.
\item {\bf{Overall:}}
how well the whole harmonization is in comprehensive consideration.
\end{itemize}
\vspace{-3pt}
A total of 102 subjects participated in the test. The ratings of 9 subjects who always gave the same score regardless of different harmonization were discarded. After the cleaning step, 93 subjects remained. 57 of them know the definition of chord progression. In addition, if the overall score contradicted the ranking result, the ratings for a melody were discarded.

\subsection{Results}
\label{sec:Result}

\subsubsection{Objective tests}
\label{sec:Objective test results}

The results are shown in Table \ref{table:objective}. Before we get into the discussion, it should be noted that none of the objective metrics mentioned in Section \ref{sec:metrics} can directly reflect the quality of the composed chords. It is difficult to quantify which metrics are better for which purposes and how accurate these metrics are. The human ear is still the best criterion for all music pieces.

\begin{table}[t]
\caption {Objective evaluation scores of different models. BGS denotes blocked Gibbs sampling and B denotes chord balancing. Higher values in CTnCTR and PCS and lower values in MCTD indicate better melody/chord harmonicity. Higher values in CHE and CC and lower values in CTD implicitly mean better interestingness of chord progression.}
\vspace{10pt}
\label{table:objective}
\centering
\begin{tabular*}{\linewidth} {@{\extracolsep{\fill}} cccc}
\toprule
    \multicolumn{1}{l} {\textbf{M/C Harmonicity}} & CTnCTR$\uparrow$ & PCS$\uparrow$ & MCTD$\downarrow$ \\
    \midrule
    \multicolumn{1}{l} {Humans} & $0.726$ & $0.515$ & $1.276$ \\
    \multicolumn{1}{l} {MTHarmonizer, $|S|=48$} & $0.804$ & $0.576$ & $1.152$ \\
    \multicolumn{1}{l} {BGS w/o B, $|S|=48$} & $0.850$ & $0.652$ & $1.087$ \\
    \multicolumn{1}{l} {BGS w/o B, $|S|=96$} & $0.853$ & $\bf{0.657}$ & $1.053$ \\
    \multicolumn{1}{l} {BGS w/ B, $|S|=48$} & $0.876$ & $0.653$ & $1.068$ \\
    \multicolumn{1}{l} {BGS w/ B, $|S|=96$} & $\bf{0.887}$ & $0.652$ & $\bf{1.052}$ \\
    \midrule
    \midrule
    \multicolumn{1}{l} {\textbf{Chord Progression}} & CHE$\uparrow$ & CC$\uparrow$ & CTD$\downarrow$ \\
    \midrule
    \multicolumn{1}{l} {Humans} & $1.266$ & $4.344$ & $0.628$ \\
    \multicolumn{1}{l} {MTHarmonizer, $|S|=48$} & $0.859$ & $3.104$ & $\bf{0.512}$ \\
    \multicolumn{1}{l} {BGS w/o B, $|S|=48$} & $0.755$ & $2.926$ & $0.521$ \\
    \multicolumn{1}{l} {BGS w/o B, $|S|=96$} & $0.726$ & $2.862$ & $0.540$ \\
    \multicolumn{1}{l} {BGS w/ B, $|S|=48$} & $1.209$ & $4.698$ & $0.690$ \\
    \multicolumn{1}{l} {BGS w/ B, $|S|=96$} & $\bf{1.280}$ & $\bf{4.900}$ & $0.730$ \\
\bottomrule
\end{tabular*}
\vspace{-10pt}
\end{table}

In terms of the melody/chord harmonicity metrics (i.e., CTnCTR, PCS, and MCTD), we can see that the human-composed pieces are worse than the automatically composed pieces. We think that this happens because human composers use many 7th, 9th, and 11th chords to produce more interesting chord progressions. However, these modern chords may degrade the performance in the melody/chord harmonicity metrics. As a result, we should not conclude that in terms of melody/chord harmonicity, the performance of human composers is worse than that of melody harmonization systems.

\begin{table}[t]
\caption {Votes and rates of the best with respect to various methods. In each of the 5 melody/chord test sets, the subject was asked to choose the best from three chord progressions.}
\vspace{10pt}
\label{table:votes}
\centering
\begin{tabular*}{0.9\linewidth} {@{\extracolsep{\fill}} ccc}
\toprule
    \multicolumn{1}{c} {\bf Method} & \bf Votes & \bf Rate of the Best \% \\
    \midrule
    \multicolumn{1}{c} {Humans} & $87$ & $31.9$  \\
    \multicolumn{1}{c} {MTHarmonizer} & $66$ & $24.2$ \\
    \multicolumn{1}{c} {\bf Proposed} & $\bf86$ & $\bf31.5$ \\
    \multicolumn{1}{c} {No preference} & $34$ & $12.5$ \\
\bottomrule
\end{tabular*}
\vspace{-15pt}
\end{table}

Comparing our model (cf. BGS w/o B, $|S|=48$) with MTharmonizer, we find that the introduction of the orderless NADE training process and blocked Gibbs sampling can indeed improve the performance in all three melody/chord harmonicity metrics. This result shows that our model can compose the chords that are highly coherent with the given melody. In addition, introducing chord balancing (i.e., class weighting) and augmenting the chord space to 96 will not only not damage the harmonicity score, but will actually increase the performance (cf. BGS w/ B, $|S|=96$). We think that this is because the more chords that can be selected and the more balanced the distribution of chord appearance, our model can better demonstrate its ability. 

For the metrics of chord progression, we can find that humans compose very interesting chord progressions. Neither MTHarmonizer nor our model (cf. BGS w/o B, $|S|=48$ or $96$) can achieve such diversity. However, with the introduction of chord balancing, our model (cf. BGS w/ B, $|S|=48$ or $96$) can compose highly diverse chord progressions. It is worth noting that our model did not get a good CTD score. However, the CTD score can only represent the distance between two chords. It will be 0 if the chord does not change throughout the chord progression. In fact, the CTD scores of our model are closer to that of human-composed chords.

\subsubsection{Subjective tests}
\label{sec:subjective test result}
\begin{figure}[!t]
\centering
\includegraphics[width=0.49\textwidth]{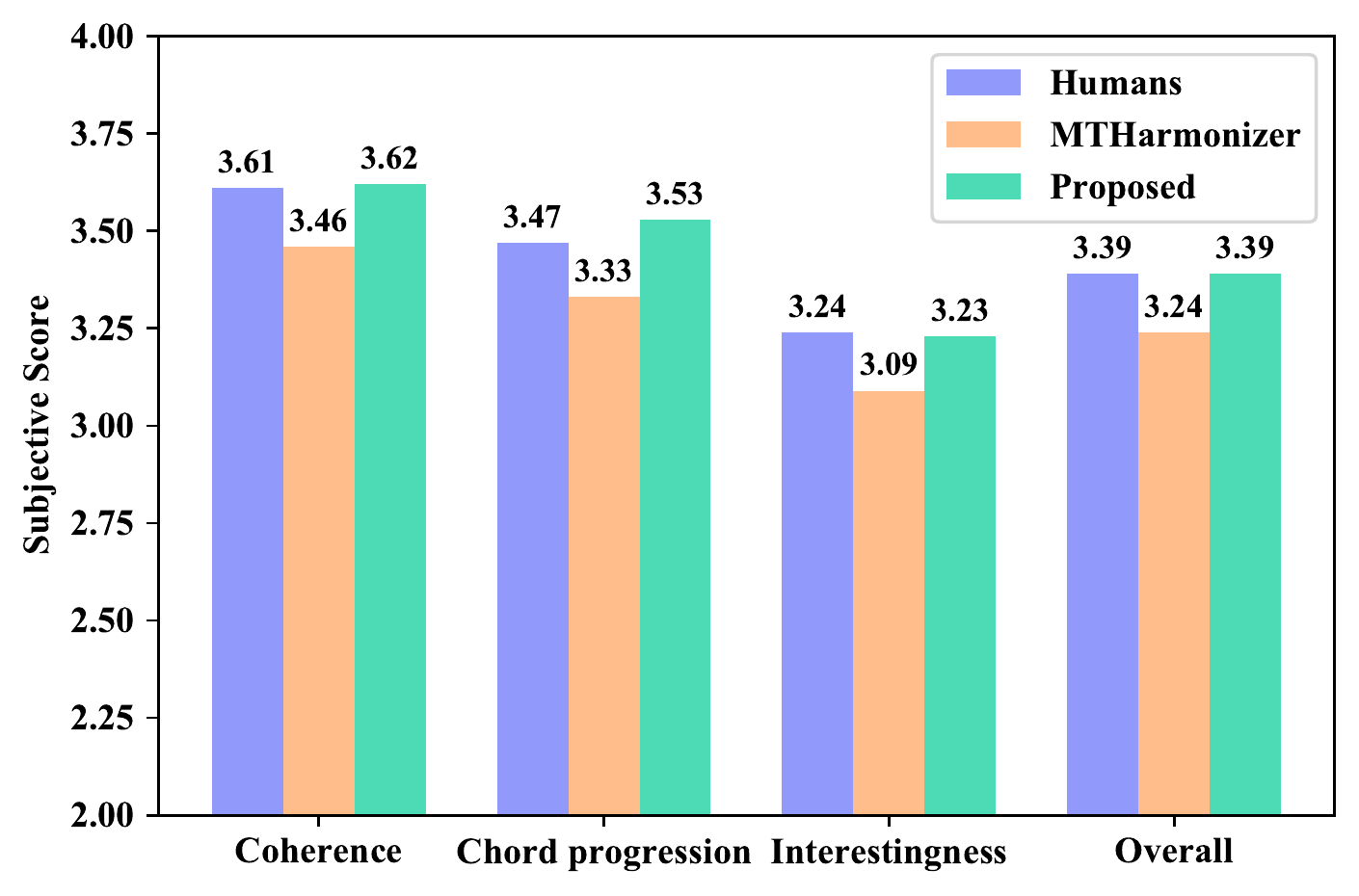}
\vspace{-20pt}
\caption{Histogram of subjective evaluation results. Our model outperforms MTHarmonizer in all aspects and even gains higher scores than human composers in chord progression.}
\label{fig:histogram}   
\vspace{-10pt}
\end{figure}

Fig. \ref{fig:histogram} shows the results of the subjective test. Our model (BGS w/ B, $|S|=96$) surpasses MTHarmonizer in all criteria and has close results to human composers. Our model has excellent performance in terms of coherence, which is consistent with the high melody/chord harmonicity in the objective evaluation. The high score also indicates that with blocked Gibbs sampling, our model can produce more reasonable and smoother chord progressions. 

The subjects were also asked to choose the best from three chord progressions of a melody. Table \ref{table:votes} shows that the music pieces composed by humans are the most favorable, but the percentage of being selected as the best is only slightly higher than that of our model. We conclude that our model can arrange music pieces very well.

Several examples are available at \url{https://chord-generation.herokuapp.com/demo}. They are randomly rendered from all 500 test samples in HLSD.

\section{Conclusions and Future Work}
\label{sec:Conclusions and Future Work}
In this paper, we proposed a framework based on the BiLSTM structure and the orderless NADE training process for automatic melody harmonization. We applied blocked Gibbs sampling in inference. In addition, we applied chord balancing and augmented the chord space to 96 to increase the diversity of chord progressions. The objective and subjective test results demonstrate that our model is better than Yeh's model and is comparable in performance to human composers. 

In our future work, we will try to make our neural model learn Roman numeral analysis, which is how humans learn how to compose chords, and one of the reasons why our proposed model cannot surpass human composers.

\bibliographystyle{IEEEtran}
\bibliography{references.bib}

\end{document}